\documentclass[12pt]{article}
\newcommand{\nc}{\newcommand}
\nc{\bib}{\bibitem}
\nc{\al}{\alpha}
\nc{\g}{\gamma}
\nc{\G}{\Gamma}
\nc{\D}{\Delta}
\nc{\eps}{\epsilon}
\nc{\la}{\lambda}
\nc{\La}{\Lambda}
\nc{\var}{\varphi}
\nc{\hn}{h^\vee}
\nc{\kn}{k^\vee}
\nc{\adg}{a^\dagger}
\nc{\bdg}{b^\dagger}
\nc{\ba}{\beta_\al}
\nc{\ga}{\g^{\al_1}}
\nc{\vpp}{{V_+}^+}
\nc{\cpp}{{C_+}^+}
\nc{\Vm}{V_{-\al^-}^{\al_1}}
\nc{\Vp}{V_{-\al^+}^{\al_1}}
\nc{\Vmb}{V_{-\beta^-}^{\al_1}}
\nc{\Gb}{\overline{G}}
\nc{\Gbc}{\overline{{\cal G}}}
\nc{\pa}{\partial}
\nc{\nn}{\nonumber \\ }
\nc{\hf}{\frac{1}{2}}         
\nc{\dz}{\frac{dz}{2\pi i}}
\nc{\fabc}{{f_{a,b}}^c}
\nc{\bin}[2]{\left (\begin{array}{c} {#1}\\ {#2} \end{array}\right )}
\nc{\ben}{\begin{equation}}
\nc{\een}{\end{equation}}
\nc{\bea}{\begin{eqnarray}}
\nc{\eea}{\end{eqnarray}}
\nc{\bra}[1]{\langle {#1}|}
\nc{\ket}[1]{|{#1}\rangle}

\nc{\C}{\mbox{\hspace{1.24mm}\rule{0.2mm}{2.5mm}\hspace{-2.7mm} C}}
\nc{\Nat}{\mbox{\hspace{.04mm}\rule{0.2mm}{2.8mm}\hspace{-1.5mm} N}}

\nc{\spa}{\hspace{.1cm},\hspace{1 cm}}
\nc{\vs}{\vspace}
\nc{\NP}[1]{Nucl.\ Phys.\ {\bf #1}}
\nc{\PL}[1]{Phys.\ Lett.\ {\bf #1}}
\nc{\CMP}[1]{Commun.\ Math.\ Phys.\ {\bf #1}}
\nc{\PR}[1]{Phys.\ Rev.\ {\bf #1}}
\nc{\PRL}[1]{Phys.\ Rev.\ Lett.\ {\bf #1}}
\nc{\PTP}[1]{Prog.\ Theor.\ Phys.\ {\bf #1}}
\nc{\PTPS}[1]{Prog.\ Theor.\ Phys.\ Suppl.\ {\bf #1}}
\nc{\MPL}[1]{Mod.\ Phys.\ Lett.\ {\bf #1}}
\nc{\IJMP}[1]{Int.\ Jour.\ Mod.\ Phys.\ {\bf #1}}
\nc{\IM}[1]{Invent.\ Math.\ {\bf #1}}
\nc{\SJNP}[1]{Sov. J. Nucl. Phys.\ {\bf #1}}
\nc{\JHEP}[1]{J.\ High\ Energy Phys.\ {\bf #1}}

\begin{document}

\topmargin -5mm
\oddsidemargin 5mm

\begin{titlepage}
\setcounter{page}{0}
\begin{flushright}
March 2000
\end{flushright}

\vs{8mm}
\begin{center}
{\Large Comments on $N=4$ Superconformal Algebras}

\vs{8mm}
{\large J{\o}rgen Rasmussen}\footnote{e-mail address: 
rasmussj@cs.uleth.ca}\\[.2cm]
{\em Physics Department, University of Lethbridge}\\
{\em Lethbridge, Alberta, Canada T1K 3M4}

\end{center}

\vs{8mm}
\centerline{{\bf{Abstract}}}
\noindent
We present a new and asymmetric $N=4$ superconformal algebra for arbitrary
central charge, thus completing our recent work on its classical analogue 
with vanishing central charge. Besides the Virasoro generator and 4
supercurrents, the algebra consists of an internal $SL(2)\otimes U(1)$
Kac-Moody algebra in addition to two spin 1/2 fermions and a bosonic
scalar. The algebra is shown to be invariant under a linear twist of the
generators, except for a unique value of the
continuous twist parameter. At this value, the invariance is broken and
the algebra collapses to the
small $N=4$ superconformal algebra. In the context of string theory, 
the asymmetric $N=4$ superconformal algebra is provided with
an explicit construction on the boundary of $AdS_3$, and is induced
by an affine $SL(2|2)$ current superalgebra residing on the world
sheet. Substituting the world sheet $SL(2|2)$ by the coset
$SL(2|2)/U(1)$ results in the small $N=4$ superconformal
algebra on the boundary of $AdS_3$.
\\[.4cm]
{\em PACS:} 11.25.Hf; 11.25.-w\\
{\em Keywords:} Superconformal field theory; 
$AdS$/CFT correspondence; string theory; free field realization

\end{titlepage}
\newpage
\renewcommand{\thefootnote}{\arabic{footnote}}
\setcounter{footnote}{0}

\section{Introduction}

Motivated by the exciting progress on the correspondence between string theory
on anti-de Sitter space ($AdS$) and conformal field theory 
\cite{Mal,GKP,Wit}, we have recently outlined an explicit construction of an
infinite dimensional class of superconformal algebras (SCAs) on the boundary 
of $AdS_3$ \cite{Ras2}. These space-time algebras are $N$ extended SCAs
induced by an affine $SL(2|N/2)$ current superalgebra residing on the world
sheet ($N$ is even). Our construction generalizes a work by Ito \cite{Ito}
in which $N=1$, 2 and 4 SCAs are studied.
Our constructions are both supersymmetric extensions
of the Giveon-Kutasov-Seiberg construction of the Virasoro
algebra \cite{GKS}. A related approach to building $N=1$, 2 and 4
SCAs for fixed central charges may be found in Ref. \cite{And}.

In the present paper we shall complete our construction of the new
and {\em asymmetric} $N=4$ SCA discovered in Ref. \cite{Ras2}, where the 
case of vanishing central charge was treated. 
Here we shall provide the algebra for generic central charge.
Besides the Virasoro generator and 4 supercurrents, the algebra
consists of an internal $SL(2)\otimes U(1)$ Kac-Moody algebra
in addition to two spin 1/2 fermions and a bosonic scalar.
Hence, it is not included in the standard classification of $N=4$ SCAs
\cite{Aetal,Sch,STV,AK,Ali}.

We shall also show that replacing the world sheet $SL(2|2)$ current
superalgebra by that of the coset $SL(2|2)/U(1)$
induces the well known small $N=4$ SCA rather than the bigger,
asymmetric $N=4$ SCA induced by $SL(2|2)$. We hope this will
clarify the incompatibility between our work and the result for
$N=4$ in Ref. \cite{Ito}; there a construction (similar to ours)
based on $SL(2|2)$ was announced to result in the small $N=4$ SCA.

The new and asymmetric $N=4$ SCA is invariant under a one-parameter
twist. The starting point for this observation is a linear modification
of the Giveon-Kutasov-Seiberg Virasoro generator.
The remaining twisted generators are obtained by requiring
the SCA to be invariant. All twisted generators are given in terms of
linear combinations of the original untwisted generators. 
It is an interesting observation that for
precisely one value of the continuous
twist parameter, the invariance is broken
and the twisting results in the small $N=4$ SCA.
It is stressed that the invariance and this collapse of the asymmetric SCA 
to the small $N=4$ SCA, are both independent of our construction and rely
solely on the defining (anti-)commutators of the two SCAs.

The rest of this paper is organized as follows.
In Section 2 we summarize briefly some of our results \cite{Ras2} on the
general construction of SCAs induced by affine $SL(2|N/2)$ current 
superalgebras.

Section 3 is devoted to $N=4$ SCAs, where the new and asymmetric algebra
is provided. The small $N=4$ is shown to be induced by the coset 
$SL(2|2)/U(1)$ current superalgebra, and the invariance of the asymmetric
SCA is discussed.

Section 4 contains concluding remarks, while details on the Lie superalgebra
$sl(2|M)$ along with explicit free field realizations of the currents,
are given in appendices A, B and C.

\section{$N$ Extended Superconformal Algebras}

\subsection{Free Field Realizations of Affine Current Superalgebras}

Associated to a Lie superalgebra ${\bf g}$ is an affine Lie superalgebra 
characterized by the central extension $k$. Associated to an affine Lie
superalgebra is an affine current superalgebra whose generators, $J_a$, 
are conformal spin one primary fields 
(with respect to the Sugawara construction)
and have the mutual operator product expansions (OPEs)
\ben
 J_a(z)J_b(w)=\frac{\kappa_{a,b}k}{(z-w)^2}+\frac{{f_{a,b}}^c J_c(w)}{z-w}
\label{JaJb}
\een
$\kappa_{a,b}$ and $\fabc$ are the Cartan-Killing form and 
structure constants, respectively, of the underlying Lie superalgebra. 
Regular terms have been omitted.
The general free field realization obtained in Ref. \cite{Ras}
is based on a pair of free ghost fields ($\ba,\g^\al$) of 
conformal weights (1,0) for every positive 
root $\al\in\D_+$ (written $\al>0$), 
and on a free scalar boson $\var_i$ for every 
Cartan index $i=1,...,r$, where $r$ is the rank of the underlying Lie
superalgebra. They satisfy the OPEs
\ben
 \ba(z)\g^{\al'}(w)=\frac{{\delta_{\al}}^{\al'}}{z-w}\spa
 \var_i(z)\var_j(w)=\kappa_{i,j}\ln(z-w)
\een
Note that in this notation, the ghost fields ($\ba,\g^\al$)
associated to {\em odd} roots $\al\in\D^1_+$ are {\em fermionic}.
($\D^1_\pm$) $\D^0_\pm$ denotes the space of positive or negative (odd) even
roots of the underlying Lie superalgebra, and $\D_\pm=\D^0_\pm\cup\D^1_\pm$. 
The corresponding energy-momentum tensor is
\ben
 T=\sum_{\al>0}\pa\g^\al\ba
  +\hf\pa\var\cdot\pa\var-\frac{1}{\sqrt{k+\hn}}\rho\cdot\pa^2\var
\label{T}
\een
$\rho$ and $\hn$ are the Weyl vector and the dual Coxeter number,
respectively, of the underlying Lie superalgebra. Normal ordering is implicit.
The generators of the affine current superalgebra
are realized according to \cite{Ras}
\ben
 J_a(z)=\sum_{\al>0}V_a^\al(\g(z))\beta_\al(z)
  +\sqrt{k+\hn}\sum_{j=1}^rP_a^j(\g(z))\pa\var_j(z)
  +J^{\mbox{\tiny{anom}}}_a(\g(z),\pa\g(z))
\een
where
\ben
 J^{\mbox{\tiny{anom}}}_a(\g(z),\pa\g(z))= 
  \left\{\begin{array}{l}0\ \ \ \mbox{for}\ \ a=i,\al>0\\   {}\\
  \sum_{\al'>0}\pa\g^{\al'}(z)F_{\al,\al'}(\g(z))\ \ \ \mbox{for}
  \ \ a=\al<0  \end{array} \right.
\een
The explicit form of $F_{\al,\al'}$ is not 
needed here but may be found in refs. \cite{Ras,Ras2}. For $\al=\al_i$ 
a simple root $F_{\al_i,\al'}$ is a constant independent of the ghost fields
$\g$:
\ben
 F_{\al_i,\al'}(\g)
  =\hf\delta_{\al_i,\al'}\left((2k+\hn)\kappa_{\al_i,-\al_i}-A_{ii}\right)
\een
$A_{ij}$ is the Cartan matrix of the underlying Lie superalgebra, and
is related to the Cartan-Killing form as 
$\kappa_{i,j}=A_{ij}\kappa_{\al_j,-\al_j}$. $V$ and $P$ are given by
\bea
  V_\al^{\al'}(\g)&=&\left[B(C(\g))\right]_\al^{\al'}\nn
  V_i^{\al'}(\g)&=&-\left[C(\g)\right]_i^{\al'}\nn
  V_{-\al}^{\al'}(\g)&=&\sum_{\al''>0}\left[e^{-C(\g)}
  \right]_{-\al}^{\al''}\left[B(-C(\g))\right]_{\al''}^{\al'}\nn
  P_\al^j(\g)&=&0\nn
  P_i^j(\g)&=&\delta_i^j\nn
  P_{-\al}^j(\g)&=&\left[e^{-C(\g)}\right]_{-\al}^j
\label{pol}
\eea
$B(u)$ is the generating function for the Bernoulli numbers $B_n$
\ben
 B(u)=\frac{u}{e^u-1}=\sum_{n\geq0}\frac{B_n}{n!}u^n
\een
whereas the matrix $C$ is defined by
\ben
 C_a^b(\g)=-\sum_{\al>0}\g^\al{f_{\al,a}}^b
\een
The formal power series (\ref{pol}) all truncate and are thus polynomials.

\subsection{Superconformal Algebra Generators}

Most Lie superalgebras with even subalgebra ${\bf g}^0=sl(2)\oplus{\bf g'}$
have the property that the embedding of $sl(2)$ in ${\bf g}$
carried by the odd generators (the set of which is denoted ${\bf g}^1$) 
is a fundamental (spin 1/2) representation\footnote{This is true
for all basic Lie superalgebras with even subalgebra  
${\bf g}^0=sl(2)\oplus{\bf g'}$ except $osp(3|2M)$ where the embedding 
is a spin 1 representation, see e.g. \cite{IMP}.}. This means that
the space of odd roots may be divided into two parts
\ben
 \D^1=\D^{1-}\cup\D^{1+}
\label{decomp}
\een
The roots $\al^\pm\in\D^{1\pm}$ are characterized by 
\ben
 \frac{\al_{sl(2)}\cdot\al^\pm}{\al_{sl(2)}^2}=\pm\hf
\een
and we have the correspondence
\ben
 \D^{1+}=\al_{sl(2)}+\D^{1-}
\een
Here $\al_{sl(2)}$ is the positive root associated to the embedded $sl(2)$.
In particular, the Lie superalgebra $sl(2|N/2)$ allows such a decomposition
of the root space. 
In the distinguished representation of $sl(2|N/2)$ discussed in Appendix A,
the embedding is associated to the simple root $\al_1$, while the
only odd simple root is $\al_2$. Furthermore, the root space enjoys the 
refined decomposition
\ben
 \D_+^1=\D_+^{1-}\cup\D_+^{1+}
\een
This refinement is not valid for all Lie superalgebras respecting
(\ref{decomp}), as $osp(1|2)$ illustrates. In the following we shall
concentrate on SCAs induced by affine $SL(2|N/2)$ current superalgebras.

Now, the Virasoro algebra is induced by the embedded $SL(2)$ and
is generated by
\bea
 L_n&=&\oint\dz{\cal L}_n(z)\nn
 {\cal L}_n&=&a_+(n)\left(\g^{\al_1}\right)^{n+1}E_{\al_1}
  +a_3(n)\left(\g^{\al_1}\right)^nH_1+a_-(n)\left(\g^{\al_1}\right)^{n-1}
  F_{\al_1}
\label{vir}
\eea
The central charge is
\ben
 c=-6k^\vee_1p_1
\een
$p_1$ is the winding number 
\ben
 p_1=\oint\dz\frac{\pa\g^{\al_1}}{\g^{\al_1}}
\label{p}
\een 
while $k^\vee_1=\kappa_{\al_1,-\al_1}k$ is the level of
the embedded $sl(2)$ or the level in the direction $\al_1$. 
The constants $a_+$, $a_3$ and $a_-$ are defined by
\ben
 a_+(n)=\hf(n-n^2),\ \ \ a_3(n)=\hf(1-n^2),\ \ \ a_-(n)=\hf(n+n^2)
\een
In Ref. \cite{Ras2} it was found that for each {\em pair} of roots 
$(\al^-,\al^+)$ we have a pair of supercurrents
of spin 3/2 with respect to (\ref{vir}):
\bea
 G_{\al^-;n+1/2}&=&\oint\dz{\cal G}_{\al^-;n+1/2}(z)\nn
 {\cal G}_{\al^-;n+1/2}&=&(n+1)(\g^{\al_1})^nE_{\al^-}-n(\g^{\al_1})^{n+1}
   E_{\al^+}
\label{G}
\eea
and
\bea
 \Gb_{-\al^-;n-1/2}&=&\oint\dz\Gbc_{-\al^-;n-1/2}(z)\nn
 \Gbc_{-\al^-;n-1/2}&=&(n-1)(\g^{\al_1})^nF_{\al^-}+n(\g^{\al_1})^{n-1}
  F_{\al^+}\nn
 &-&n(n-1)(\g^{\al_1})^{n-2}\Vm\left((\g^{\al_1})^2E_{\al_1}+\g^{\al_1}H_1
  -F_{\al_1}\right)\nn
 &+&n(n-1)(\g^{\al_1})^{n-2}\sum_{\nu,\sigma>0}
  \left((\g^{\al_1})^2V_{\al_1}^\nu+\g^{\al_1}V_1^\nu
  -V_{-\al_1}^\nu\right)\pa_\nu\pa_\sigma\Vm\pa\g^\sigma
\label{Gb}
\eea
Their anti-commutators are
\ben
 \left\{G_{\al^-;n+1/2},G_{\beta^-;m+1/2}\right\}=0
\label{GG}
\een
and
\ben
 \left\{G_{\al^-;n+1/2},\Gb_{-\beta^-;m-1/2}\right\}
  =\delta_{\al^-,\beta^-}L_{n+m}+(n-m+1)K_{\al^-;-\beta^-;n+m}
  +\frac{1}{6}cn(n+1)\delta_{n+m,0}\delta_{\al^-,\beta^-}
\label{GGb}
\een
where the primary spin 1 (with respect to (\ref{vir}))
current $K$ is defined by
\bea
 K_{\al^-;-\beta^-;n}&=&\oint\dz{\cal K}_{\al^-;-\beta^-;n}(z)\nn
 {\cal K}_{\al^-;-\beta^-;n}&=&
  n(\g^{\al_1})^{n-1}\Vmb\left(\g^{\al_1}E_{\al^+}-E_{\al^-}\right)
  -(\g^{\al_1})^n{f_{\al^-,-\beta^-}}^cJ_c\nn
 &+&\hf\delta_{\al^-,\beta^-}(\g^{\al_1})^{n-1}\left(n\left((\g^{\al_1})^2
  E_{\al_1}+\g^{\al_1}H_1-F_{\al_1}\right)-
  \g^{\al_1}H_1\right)\nn
 &+&n(\g^{\al_1})^{n-1}\sum_{\nu,\sigma>0}
  \left(\g^{\al_1}V_{\al^+}^\nu-V_{\al^-}^\nu\right)
  \pa_\nu\pa_\sigma\Vmb\pa\g^\sigma
\label{K}
\eea
It was also shown in Ref. \cite{Ras2}, 
that the remaining anti-commutators are generally non-vanishing:
\ben
 \left\{\Gb_{-\al^-;n-1/2},\Gb_{-\beta^-;m-1/2}\right\}\neq0,\ \ \
  \mbox{for}\ \ \al^-\neq\beta^-,\ n\neq m,\ n+m\neq1
\label{GbGb}
\een
This asymmetric property of the SCAs induced by $SL(2|N/2)$ will be
illustrated in the following where we consider the case $N=4$.

\section{$N=4$ Superconformal Algebras}

Here we shall specialize the general considerations on $SL(2|N/2)$ in 
Section 2 to the case $N=4$. The resulting SCA is of a new and asymmetric
form. Its classical and centerless analogue has recently been 
obtained in Ref. \cite{Ras2}. Below we shall provide the full SCA for
arbitrary central charge. We shall furthermore show that 
substituting the original $SL(2|2)$ by the coset $SL(2|2)/U(1)$
reduces the asymmetric $N=4$ SCA to the well known small $N=4$ SCA.
An invariance of the asymmetric $N=4$ SCA is also presented.

\subsection{Asymmetric $N=4$ SCA from $SL(2|2)$}

In the distinguished representation discussed in Appendix A,
the Cartan matrix and the Cartan-Killing form of the Lie superalgebra
$SL(2|2)$ are
\ben
 A_{ij}=\left(\begin{array}{rrr} 2&-1&0\\  -1&0&1\\  0&-1&2 \end{array}
  \right),\ \ \ 
 \kappa_{i,j}=\left(\begin{array}{rrr}2&-1&0\\ -1&0&-1\\ 0&-1&-2\end{array}
  \right)
\label{Cartan}
\een
The dual Coxeter number is $\hn=0$ while the number of supercurrents
is $2|\D_+^{1-}|=4$. Accordingly, there are a priori 4 generators, $K$, 
of the internal Kac-Moody algebra.
Using these facts, with the explicit realizations of the 
Virasoro generators (\ref{vir}), the supercurrents (\ref{G}) 
and (\ref{Gb}), and the affine currents (\ref{K}) (given in
Appendix C), one may work out the entire SCA.
We find that closure is ensured by the following set of generators
\bea
 \mbox{Virasoro\ generator}\ \ &L&\ \ h=2\nn
 \mbox{supercurrents}\ \ &G_{\al_2},\ G_{\al_2+\al_3},\ \Gb_{-\al_2},\ 
  \Gb_{-(\al_2+\al_3)}&\ \ h=3/2\nn
 \mbox{affine}\ SL(2)\ \ &\tilde{E}=K_{\al_2+\al_3;-\al_2},&\nn 
 &\tilde{H}=K_{\al_2+\al_3;-(\al_2+\al_3)}-K_{\al_{2};-\al_2},&\nn
  &\tilde{F}=K_{\al_{2};-(\al_2+\al_3)}&\ \ h=1\nn
 \mbox{affine} \ U(1)\ \ &U=K_{\al_2+\al_3;-(\al_2+\al_3)}
  +K_{\al_{2};-\al_2}&\ \ h=1\nn
 \mbox{fermions}\ \ &\phi_{-\al_2},\ \phi_{-(\al_2+\al_3)}&\ \ h=1/2\nn
 \mbox{scalar}\ \ &S&\ \ h=0
\label{agen}
\eea
and that the non-trivial (anti-)commutators are
\bea
 \left[L_n,L_m\right]&=&(n-m)L_{n+m}+\frac{c}{12}(n^3-n)\delta_{n+m,0}\nn
 \left[L_n,A_m\right]&=&((h(A)-1)n-m)A_{n+m}\nn
 \left\{G_{\al^-;n+1/2},G_{\beta^-;m+1/2}\right\}&=&
  \left\{\Gb_{-\al^-;n-1/2},\Gb_{-\al^-;m-1/2}\right\}=0\nn
 \left\{G_{\al^-;n+1/2},\Gb_{-\beta^-;m-1/2}\right\}&=&\delta_{\al^-,\beta^-}
  L_{n+m}+(n-m+1)K_{\al^-;-\beta^-;n+m}\nn
 &+&\frac{1}{6}cn(n+1)\delta_{n+m,0}\delta_{\al^-,\beta^-}\nn
 \left\{\Gb_{-\al_2;n-1/2},\Gb_{-(\al_2+\al_3);m-1/2}\right\}&=&(n-m)
  (n+m-1)S_{n+m-1}\nn
 \left[\tilde{H}_n,\tilde{E}_m\right]&=&2\tilde{E}_{n+m},\ \ 
 \left[\tilde{H}_n,\tilde{F}_m\right]=-2\tilde{F}_{n+m}\nn
 \left[\tilde{E}_n,\tilde{F}_m\right]&=&\tilde{H}_{n+m}
  +\frac{1}{6}cn\delta_{n+m,0},\ \ \left[\tilde{H}_n,\tilde{H}_m\right]
  =\frac{1}{3}cn\delta_{n+m,0}\nn
 \left[\tilde{E}_n,G_{\al_2;m+1/2}\right]&=&G_{\al_2+\al_3;n+m+1/2},\ \ 
 \left[\tilde{F}_n,G_{\al_2+\al_3;m+1/2}\right]=G_{\al_2;n+m+1/2}\nn
 \left[\tilde{H}_n,G_{\al_2;m+1/2}\right]&=&-G_{\al_2;n+m+1/2}\nn
 \left[\tilde{H}_n,G_{\al_2+\al_3;m+1/2}\right]&=&G_{\al_2+\al_3;n+m+1/2}\nn
 \left[\tilde{E}_n,\Gb_{-(\al_2+\al_3);m-1/2}\right]&=&
  -\Gb_{-\al_{2};n+m-1/2}-n\phi_{-\al_2;n+m-1/2}\nn
 \left[\tilde{H}_n,\Gb_{-\al_{2};m-1/2}\right]&=&
  \Gb_{-\al_{2};n+m-1/2}+n\phi_{-\al_2;n+m-1/2}\nn
 \left[\tilde{H}_n,\Gb_{-(\al_2+\al_3);m-1/2}\right]&=&
  -\Gb_{-(\al_2+\al_3);n+m-1/2}-n\phi_{-(\al_2+\al_3);n+m-1/2}\nn
 \left[\tilde{F}_n,\Gb_{-\al_{2};m-1/2}\right]&=&
  -\Gb_{-(\al_2+\al_3);n+m-1/2}-n\phi_{-(\al_2+\al_3);n+m-1/2}\nn
 \left[U_n,\Gb_{-\al^-;m-1/2}\right]&=&n\phi_{-\al^-;n+m-1/2}\nn
  \left[S_{n},G_{\al_2;m+1/2}\right]&=&\phi_{-(\al_2+\al_3);n+m+1/2}\nn
 \left[S_{n},G_{\al_2+\al_3;m+1/2}\right]&=&-\phi_{-\al_{2};n+m+1/2}\nn 
 \left\{G_{\al_2;n+1/2},\phi_{-\al_2;m-1/2}\right\}&=&U_{n+m},\ \
 \left\{G_{\al_2+\al_3;n+1/2},\phi_{-(\al_2+\al_3);m-1/2}\right\}=U_{n+m}\nn
 \left\{\Gb_{-\al_2;n-1/2},\phi_{-(\al_2+\al_3);m-1/2}\right\}&=&
  (n+m-1)S_{n+m-1}\nn
 \left\{\Gb_{-(\al_2+\al_3);n-1/2},\phi_{-\al_{2};m-1/2}\right\}&=&
  -(n+m-1)S_{n+m-1}\nn 
 \left[\tilde{E}_n,\phi_{-(\al_2+\al_3);m-1/2}\right]
  &=&-\phi_{-\al_{2};n+m-1/2}\nn
 \left[\tilde{F}_n,\phi_{-\al_{2};m-1/2}\right]
  &=&-\phi_{-(\al_2+\al_3);n+m-1/2}\nn
 \left[\tilde{H}_n,\phi_{-\al_2;m-1/2}\right]&=&\phi_{-\al_2;n+m-1/2}\nn
 \left[\tilde{H}_n,\phi_{-(\al_2+\al_3);m-1/2}\right]
  &=&-\phi_{-(\al_2+\al_3);n+m-1/2}\nn
 \left[U_n,U_m\right]&=&\left[S_n,S_m\right]=0
\label{N4a}
\eea
$A_m$ denotes any of the 11 generators listed in (\ref{agen})
different from the Virasoro generator.
In the derivation we have used that integrating a total derivative
gives zero. In particular, we find
\bea
 &&\left\{\Gb_{-\al_2;n-1/2},\Gb_{-(\al_2+\al_3);m-1/2}\right\}\nn
 &=&(n-m)(n+m-1)\left(S_{n+m-1}
  +\oint\dz\frac{\pa}{\pa z}\left[\g^{n+m-2}(z)
  V_{-\al_2}^{\al_1}(\g(z))V_{-(\al_2+\al_3)}^{\al_1}(\g(z))\right]
  \right)\nn
 &=&(n-m)(n+m-1)S_{n+m-1}
\eea
The polynomials $V_{-\al_2}^{\al_1}$ and $V_{-(\al_2+\al_3)}^{\al_1}$
are given (\ref{V}) in Appendix C.

It is observed that this $N=4$ SCA (\ref{N4a}) is a new and asymmetric one
not contained in the standard classification of $N=4$ SCAs
\cite{Aetal,Sch,STV,AK,Ali}. Besides being asymmetric in the way
the $G$ and $\Gb$ supercurrents are treated, it involves the unfamiliar
number {\em two} of spin 1/2 fermions. We recall that the small $N=4$ SCA 
does not contain any spin 1/2 fermions, whereas the big $N=4$ SCAs are 
characterized by containing 4 such generators.

\subsection{Small $N=4$ SCA from Coset $SL(2|2)/U(1)$}

Among the Lie superalgebras $sl(2|M)$, $M\geq1$,
$sl(2|2)$ is the only one having a
non-trivial center. This may be seen easily
by considering the associated Cartan matrices; they are all invertible
except when $M=2$. By simple inspection of the Cartan matrix for $sl(2|2)$
(\ref{Cartan}) it follows that the Lie superalgebra element 
\ben
 H_{u(1)}=H_1+2H_2-H_3
\label{hu1}
\een
generates the center $u(1)$ of $sl(2|2)$.
The coset algebra $sl(2|2)/u(1)$ may therefore be realized 
straightforwardly in terms
of the generators of the original $sl(2|2)$. All that one has to do 
is to invoke the vanishing of the generator of the center (\ref{hu1}), 
and otherwise make no changes. It should be noted that the root space
of the coset algebra is identified with the root space of $sl(2|2)$.
In particular, the sets of simple roots are identical, despite the
fact that the rank of the coset algebra is one smaller than the
rank of $sl(2|2)$, the latter being $r=3$.

The associated affine $SL(2|2)/U(1)$ current superalgebra is equally
simple to realize. By construction, its Virasoro generator
$T_{SL(2|2)/U(1)}=T_{SL(2|2)}-T_{U(1)}$ has vanishing OPE with the
current
\bea
 H_{U(1)}(z)&=&H_1(z)+2H_2(z)-H_3(z)\nn
 &=&\sqrt{k}(\pa\var_1(z)+2\pa\var_2(z)-\pa\var_3(z))
\label{hu1z}
\eea
while $T_{U(1)}$ has vanishing OPEs with all other affine currents.
Thus, it makes sense from the conformal field theory point of view
to put the central current (\ref{hu1z})
equal to zero without modifying the Virasoro generator of the original
$SL(2|2)$ current superalgebra, but by imposing the (coset) condition
\ben
 \pa\var_1+2\pa\var_2-\pa\var_3\equiv0
\label{ffrcoset}
\een
In this way the coset current superalgebra has the same free field 
realization as the original current superalgebra, though subject to 
the coset condition (\ref{ffrcoset}). Thus, at 
the level of the free field realization the coset condition 
(\ref{ffrcoset}) reflects modding out the $U(1)$ center of $SL(2|2)$. 

{}From the explicit realization of the generators of the
asymmetric $N=4$ SCA induced by $SL(2|2)$ (see Appendix C)
it follows that imposing the coset 
condition (\ref{ffrcoset}) has fundamental consequences for the resulting
$N=4$ SCA. Even the number of generators is reduced as the affine
$U(1)$ subalgebra generated by $U$ (\ref{Ureal}), 
the 2 spin 1/2 fermions $\phi$ (\ref{phireal}), 
and the bosonic scalar $S$ (\ref{Sreal}) all vanish identically 
\ben
 U_n\equiv\phi_{-\al_2;n-1/2}\equiv\phi_{-(\al_2+\al_3);n-1/2}\equiv
 S_n\equiv0
\een
We conclude that the $N=4$ SCA induced by the affine $SL(2|2)/U(1)$
current superalgebra is generated by  
\bea
 \mbox{Virasoro\ generator}\ \ &L&\ \ h=2\nn
 \mbox{supercurrents}\ \ &G_{\al_2},\ G_{\al_2+\al_3},\ \Gb_{-\al_2},\ 
  \Gb_{-(\al_2+\al_3)}&\ \ h=3/2\nn
 \mbox{affine}\ SL(2)\ \ &\tilde{E}=K_{\al_2+\al_3;-\al_2},&\nn 
 &\tilde{H}=K_{\al_2+\al_3;-(\al_2+\al_3)}-K_{\al_{2};-\al_2},&\nn
  &\tilde{F}=K_{\al_{2};-(\al_2+\al_3)}&\ \ h=1
\label{smallgen}
\eea
where $U=K_{\al_2+\al_3;-(\al_2+\al_3)}+K_{\al_{2};-\al_2}=0$.
The non-trivial (anti-)commutators are
\bea
 \left[L_n,L_m\right]&=&(n-m)L_{n+m}+\frac{c}{12}(n^3-n)\delta_{n+m,0}\nn
 \left[L_n,A_m\right]&=&((h(A)-1)n-m)A_{n+m}\nn
 \left\{G_{\al^-;n+1/2},G_{\beta^-;m+1/2}\right\}&=&
  \left\{\Gb_{-\al^-;n-1/2},\Gb_{-\beta^-;m-1/2}\right\}=0\nn
  \left\{G_{\al^-;n+1/2},\Gb_{-\beta^-;m-1/2}\right\}&=&\delta_{\al^-,\beta^-}
   L_{n+m}+(n-m+1)K_{\al^-;-\beta^-;n+m}\nn
 &+&\frac{1}{6}cn(n+1)\delta_{n+m,0}\delta_{\al^-,\beta^-}\nn
 \left[\tilde{E}_n,G_{\al_2;m+1/2}\right]&=&G_{\al_2+\al_3;n+m+1/2},\ \ 
 \left[\tilde{F}_n,G_{\al_2+\al_3;m+1/2}\right]=G_{\al_2;n+m+1/2}\nn
 \left[\tilde{H}_n,G_{\al_2;m+1/2}\right]&=&-G_{\al_2;n+m+1/2},\ \
 \left[\tilde{H}_n,G_{\al_2+\al_3;m+1/2}\right]=G_{\al_2+\al_3;n+m+1/2}\nn
 \left[\tilde{E}_n,\Gb_{-(\al_2+\al_3);m-1/2}\right]&=&
  -\Gb_{-\al_{2};n+m-1/2}\nn
 \left[\tilde{F}_n,\Gb_{-\al_{2};m-1/2}\right]&=&
  -\Gb_{-(\al_2+\al_3);n+m-1/2}\nn
 \left[\tilde{H}_n,\Gb_{-\al_{2};m-1/2}\right]&=&
  \Gb_{-\al_{2};n+m-1/2}\nn
 \left[\tilde{H}_n,\Gb_{-(\al_2+\al_3);m-1/2}\right]&=&
  -\Gb_{-(\al_2+\al_3);n+m-1/2}\nn
 \left[\tilde{H}_n,\tilde{E}_m\right]&=&2\tilde{E}_{n+m},\ \ 
 \left[\tilde{H}_n,\tilde{F}_m\right]=-2\tilde{F}_{n+m}\nn
 \left[\tilde{E}_n,\tilde{F}_m\right]&=&\tilde{H}_{n+m}+\frac{1}{6}
  cn\delta_{n+m,0},\ \ 
 \left[\tilde{H}_n,\tilde{H}_m\right]=\frac{1}{3}cn\delta_{n+m,0}
\label{N4small}
\eea
$A_m$ denotes any of the 7 generators listed in (\ref{smallgen})
different from the Virasoro generator.
The SCA (\ref{N4small}) is recognized as the well known small $N=4$ SCA
thus proving our assertion.

\subsection{An Invariance and a Reduction}

It turns out that the new and asymmetric $N=4$ SCA possesses an invariance
which may be revealed by considering the consequences of 
modifying the Virasoro generators according to
\ben
 L_n^\la=L_n+\la(n+1)U_n
\een
with $\la$ arbitrary. The Virasoro algebra is easily seen to be generated
with unchanged central charge
\ben
 c^\la=c
\een
We find that the asymmetric $N=4$ SCA (\ref{N4a}) is invariant under
the following one-parameter twist of its generators
\bea
 L_n^\la&=&L_n+\la(n+1)U_n\nn
 G^\la_{\al^-;n+1/2}&=&G_{\al^-;n+1/2}\nn
 \Gb^\la_{-\al^-;n-1/2}&=&\Gb_{-\al^-;n-1/2}+2\la n\phi_{-\al^-;n-1/2}\nn
 K_{\al^-;-\beta^-;n}^\la&=&K_{\al^-;-\beta^-;n}-\la\delta_{\al^-,\beta^-}U_n
  \nn
 \phi^\la_{-\al^-;n-1/2}&=&(1-2\la)\phi_{-\al^-;n-1/2}\nn
 S_n^\la&=&(1-2\la)S_n
\label{twist}
\eea
In particular, the 11 generators besides $L^\la$ are all primary of
unchanged weights
\ben
 [L_n^\la,A_m^\la]=((h(A)-1)n-m)A_{n+m}^\la
\een
It should be mentioned that one may obtain the twisted supercurrents by 
following the exact same procedure which leads to the construction of the
untwisted supercurrents, see Ref. \cite{Ras2}. In particular, 
we have the relations
\bea
 \left[L_n^\la,\oint\dz E_{\al^-}\right]&=&\hf(n-1)G_{\al^-;n+1/2}^\la\nn
 \left[L_n^\la,\oint\dz F_{\al^-}\right]&=&-\hf(n+1)\Gb_{-\al^-;n-1/2}^\la
\eea
The remaining twisted generators may of course be found by working out the 
relevant (anti-)commutators of twisted generators already obtained, and
requiring the algebra to be invariant.
For the $K$ generators, let us write out the result of the twisting
(\ref{twist})
\bea
 \tilde{E}_n^\la=\tilde{E}_n,&\tilde{H}_n^\la=\tilde{H}_n,& 
  \tilde{F}_n^\la=\tilde{F}_n\nn
 &U^\la_n=(1-2\la)U_n&
\eea
We observe that for the unique value 
\ben
 \la=1/2
\een
twisting (\ref{twist})
is not an invariance but rather a reduction. The 4 generators
$U^{\la=1/2}$, $\phi^{\la=1/2}_{-\al^-}$ and $S^{\la=1/2}$ all vanish
identically, and the resulting SCA is instead the small $N=4$ SCA.

So far we have not addressed the question of BRST invariance of our
construction of the $N=4$ SCAs. 
As pointed out in Ref. \cite{GKS}, BRST invariance of the construction
of the space-time conformal algebra based on a world sheet $SL(2)$ current
algebra, is equivalent to requiring the Virasoro
generators to be primary fields of weight one with respect to the world sheet
energy-momentum tensor, ensuring that the integrated fields commute with
the world sheet Virasoro algebra. This carries over to the superconformal 
case where all the generators are required to be integrated spin one
primary fields with respect to the world sheet Virasoro algebra. 
In Ref. \cite{Ras2} we have shown that all the generators of
the $SL(2|2)$ induced SCA are indeed BRST invariant.
Since the twisted generators (\ref{twist}) are 
linear combinations of those fields, they are themselves BRST invariant.
The generators of the small $N=4$ SCA induced by the coset
$SL(2|2)/U(1)$ in Section 3.2, are also BRST invariant. This follows 
immediately, as the construction is based on the same free field
realization as the original $SL(2|2)$, the only (and in this respect
trivial) difference being the coset condition (\ref{ffrcoset}). 

\section{Conclusion}

We believe that the general construction of SCAs outlined in Ref. \cite{Ras2}
(and indicated in Ref. \cite{Ito}) is interesting from a mathematical as
well as from a string theoretical point of view. A mathematical or
conformal field theoretical interest lies in the fact that besides
providing new realizations of well known SCAs, the construction also
produces whole new classes of SCAs. An additional virtue is that the SCAs
are realized explicitly. The asymmetric $N=4$ SCA discussed in the present
paper is an example of such a new SCA. There are also convincing indications
that new bosonic extensions of the Virasoro algebra may be
obtained using a modification of the construction.
This will be the subject of a forthcoming publication.

The construction is interesting from the string theoretical point of view,
as it produces the unique boundary SCA associated to a string theory
on $AdS_3$ with a certain affine Lie supergroup symmetry. As already
pointed out, this pertains to Lie supergroups with $SL(2)\otimes G'$
decomposable bosonic subgroups. Due to the recently discovered
$AdS$/CFT correspondence,
the question of determining which superconformal field theory is associated 
to which string theory has become increasingly relevant. 
We hope that our work will add to the understanding of this.\\[.4cm]
{\bf Acknowledgment}\\[.2cm]
The author thanks Spenta Wadia and Mark Walton for comments. 
The author is also
grateful to The Niels Bohr Institute, where part of this work was done,
for its kind hospitality. He is supported in part by NSERC of Canada.

\appendix
\section{Lie Superalgebra $sl(2|M)$}

The root space of the Lie superalgebra $sl(2|M)$ in the distinguished
representation may be realized in terms
of an orthonormal two-dimensional basis $\left\{\eps_1,\eps_2\right\}$
and an orthonormal $M$-dimensional basis $\left\{\delta_u\right\}_{u=1,...,M}$
with metrics
\ben
 \eps_{\iota}\cdot\eps_{\iota'}=\delta_{\iota,\iota'}
  \spa\delta_u\cdot\delta_{u'}=-\delta_{u,u'}\spa\eps_\iota\cdot\delta_u=0
\een
The $\hf(M+1)(M+2)$ positive roots are then represented as
\bea
 \D^0_+&=&\left\{\eps_1-\eps_2\right\}\cup\left\{\delta_u-\delta_v\ |\ 
  u<v\right\}\nn
 \D^{1+}_+&=&\left\{\eps_1-\delta_u\ |\ u=1,...,M\right\}\nn
 \D^{1-}_+&=&\left\{\eps_2-\delta_u\ |\ u=1,...,M\right\}
\eea
where the $M+1$ simple roots $\al_i$ are 
\bea
 \al_1&=&\eps_1-\eps_2\nn
 \al_2&=&\eps_2-\delta_1\nn
 \al_{u+2}&=&\delta_u-\delta_{u+1}
\eea
The associated ladder operators $E_\al$ and $F_\al$, 
and the Cartan generators $H_i$ admit 
a standard oscillator realization (see e.g. \cite{Tan})
\bea
 E_{\eps_1-\eps_2}=\adg_1a_2\spa &E_{\eps_\iota-\delta_u}=\adg_\iota b_u\spa
  &E_{\delta_u-\delta_v}=\bdg_ub_v\nn
 F_{\eps_1-\eps_2}=\adg_2a_1\spa &F_{\eps_\iota-\delta_u}=\bdg_ua_\iota\spa
  &F_{\delta_u-\delta_v}=\bdg_vb_u\nn
 H_1=\adg_1a_1-\adg_2a_2\spa &H_2=\adg_2a_2+\bdg_1b_1\spa
  &H_{u+2}=\bdg_ub_u-\bdg_{u+1}b_{u+1}
\label{osc}
\eea
where $a^{(\dagger)}_\iota$ and $b^{(\dagger)}_u$ are fermionic and bosonic
oscillators, respectively, satisfying
\ben
 \left\{a_\iota,\adg_{\iota'}\right\}=\delta_{\iota,\iota'}
  \spa \left[b_u,\bdg_v\right]
  =\delta_{u,v}\spa \left[b^{(\dagger)}_u,a^{(\dagger)}_\iota\right]=0
\een

\section{Free Field Realization of Affine $SL(2|2)$ Current Superalgebra}

In this appendix we shall provide the explicit free field realization
of the affine $SL(2|2)$ current superalgebra that is discussed in
Section 2.
Recall that the only bosonic ghost fields are $\ga,\g^{\al_3},\beta_{\al_1},
\beta_{\al_3}$.
Let us introduce the abbreviations $\g^1,\g^{23},\beta_{123},...$ for
the ghost fields $\ga,\g^{\al_2+\al_3},\beta_{\al_1+\al_2+\al_3},...$.
The free field realization is
\bea
 E_{\al_1}&=&\beta_1-\hf\g^2\beta_{12}+\hf\left(\frac{1}{6}\g^2\g^3
  -\g^{23}\right)\beta_{123}\nn
 E_{\al_2}&=&\beta_2+\hf\g^1\beta_{12}-\hf\g^3\beta_{23}-\frac{1}{6}
  \g^1\g^3\beta_{123}\nn
 E_{\al_3}&=&\beta_3+\hf\g^2\beta_{23}+\hf\left(\frac{1}{6}\g^1\g^2+\g^{12}
  \right)\beta_{123}\nn
 E_{\al_1+\al_2}&=&\beta_{12}-\hf\g^3\beta_{123}\nn
 E_{\al_2+\al_3}&=&\beta_{23}+\hf\g^1\beta_{123}\nn
 E_{\al_1+\al_2+\al_3}&=&\beta_{123}\nn
 H_1&=&-2\g^1\beta_1+\g^2\beta_2-\g^{12}\beta_{12}+\g^{23}\beta_{23}
  -\g^{123}\beta_{123}+\sqrt{k}\pa\var_1\nn
 H_2&=&\g^1\beta_1-\g^3\beta_3+\g^{12}\beta_{12}-\g^{23}\beta_{23}
  +\sqrt{k}\pa\var_2\nn
 H_3&=&\g^2\beta_2-2\g^3\beta_3+\g^{12}\beta_{12}-\g^{23}\beta_{23}
  -\g^{123}\beta_{123}+\sqrt{k}\pa\var_3\nn
 F_{\al_1}&=&-(\g^1)^2\beta_{1}+\left(\hf\g^1\g^{2}-\g^{12}
  \right)\beta_2-\hf\g^1\left(\hf\g^1\g^2+\g^{12}\right)\beta_{12}\nn
 &+&\left(\frac{1}{12}\g^1\g^2\g^3+\hf\g^1\g^{23}-\g^{123}\right)
  \beta_{23}-\hf\g^1\left(\hf\g^1\g^{23}+\frac{1}{6}\g^{12}\g^3
  +\g^{123}\right)\beta_{123}\nn
 &+&\g^1\sqrt{k}\pa\var_1+(k-1)\pa\g^1\nn
 F_{\al_2}&=&\left(\hf\g^1\g^2+\g^{12}\right)\beta_1-\left(\hf\g^2\g^3
  -\g^{23}\right)\beta_3+\hf\g^2\g^{12}\beta_{12}\nn
 &-&\hf\g^2\g^{23}\beta_{23}+\frac{1}{6}\g^2\left(\g^1\g^{23}+\g^{12}\g^3
  \right)\beta_{123}+\g^2\sqrt{k}\pa\var_2+k\pa\g^2\nn
 F_{\al_3}&=&\left(\hf\g^2\g^3+\g^{23}\right)\beta_2-(\g^3)^2\beta_3
  -\left(\frac{1}{12}\g^1\g^2\g^3-\hf\g^{12}\g^3-\g^{123}\right)
  \beta_{12}\nn
 &+&\hf\g^3\left(\hf\g^2\g^3-\g^{23}\right)\beta_{23}+\hf\g^3\left(
  \frac{1}{6}\g^1\g^{23}+\hf\g^{12}\g^3-\g^{123}\right)\beta_{123}\nn
 &+&\g^3\sqrt{k}\pa\var_3-(k+1)\pa\g^3\nn
 F_{\al_1+\al_2}&=&-\g^1\left(\hf\g^1\g^2+\g^{12}\right)\beta_1-\g^2\g^{12}
  \beta_2\nn
 &+&\left(\frac{1}{6}\g^1\g^2\g^3-\hf\g^1\g^{23}-\hf\g^{12}\g^3+\g^{123}
  \right)\beta_3\nn
 &+&\hf\g^2\left(\hf\g^1\g^{23}-\hf\g^{12}\g^3-\g^{123}\right)\beta_{23}\nn
 &-&\hf\left(\frac{5}{12}(\g^1)^2\g^2\g^{23}+\frac{1}{4}\g^1\g^2\g^{12}\g^3
  +\frac{1}{6}\g^1\g^2\g^{123}+\hf\g^1\g^{12}\g^{23}+\g^{12}\g^{123}
  \right)\beta_{123}\nn
 &+&\left(\hf\g^1\g^2+\g^{12}\right)\sqrt{k}\pa\var_1-\left(\hf\g^1\g^2
  -\g^{12}\right)\sqrt{k}\pa\var_2\nn
 &+&\frac{6k-11}{12}\g^2\pa\g^1-\frac{3k+1}{6}
  \g^1\pa\g^2+\left(k-1/2\right)\pa\g^{12}\nn
 F_{\al_2+\al_3}&=&\left(\frac{1}{6}\g^1\g^2\g^3+\hf\g^1\g^{23}+\hf
  \g^{12}\g^3+\g^{123}\right)\beta_1+\g^2\g^{23}\beta_2\nn
 &-&\g^3\left(\hf\g^2\g^3-\g^{23}\right)\beta_3-\hf\g^2\left(
  \hf\g^1\g^{23}-\hf\g^{12}\g^3-\g^{123}\right)\beta_{12}\nn
 &+&\hf\left(\frac{1}{4}\g^1\g^2\g^{23}\g^3+\frac{5}{12}\g^2\g^{12}
  (\g^3)^2-\frac{1}{6}\g^2\g^{123}\g^3+\hf\g^{12}\g^{23}\g^3+\g^{23}\g^{123}
  \right)\beta_{123}\nn
 &+&\left(\hf\g^2\g^3+\g^{23}\right)\sqrt{k}\pa\var_2+\left(\hf\g^2\g^3
  -\g^{23}\right)\sqrt{k}\pa\var_3\nn
 &+&\frac{3k-1}{6}\g^3\pa\g^2-\frac{6k+11}{12}
  \g^2\pa\g^3+(k+1/2)\pa\g^{23}\nn
 F_{\al_1+\al_2+\al_3}&=&-\g^1\left(\frac{1}{6}\g^1\g^2\g^3+\hf\g^1\g^{23}+
  \hf\g^{12}\g^3+\g^{123}\right)\beta_1\nn
 &-&\left(\hf\g^1\g^2\g^{23}+\hf\g^2\g^{12}
  \g^3-\g^{12}\g^{23}\right)\beta_2\nn
 &+&\g^3\left(\frac{1}{6}\g^1\g^2\g^3-\hf\g^1\g^{23}-\hf\g^{12}\g^3
  +\g^{123}\right)\beta_3\nn
 &+&\left(\frac{1}{4}(\g^1)^2\g^2\g^{23}+\frac{1}{12}\g^1\g^2\g^{12}\g^3
  +\g^{12}\g^{123}\right)\beta_{12}\nn
 &-&\left(\frac{1}{4}\g^2\g^{12}(\g^3)^2+\frac{1}{12}\g^1\g^2\g^{23}\g^3
  +\g^{23}\g^{123}\right)\beta_{23}\nn
 &-&\frac{1}{6}\g^1\g^3\left(\hf\g^1\g^2\g^{23}+\hf\g^2\g^{12}\g^3
  +\g^{12}\g^{23}\right)\beta_{123}\nn
 &+&\left(\frac{1}{6}\g^1\g^2\g^3+\hf\g^1\g^{23}+\hf\g^{12}\g^3
  +\g^{123}\right)\sqrt{k}\pa\var_1\nn
 &-&\left(\frac{1}{3}\g^1\g^2\g^3+\hf\g^1\g^{23}-\hf\g^{12}\g^3-\g^{123}
  \right)\sqrt{k}\pa\var_2\nn
 &-&\left(\frac{1}{6}\g^1\g^2\g^3-\hf\g^1\g^{23}-\hf\g^{12}\g^3
  +\g^{123}\right)\sqrt{k}\pa\var_3\nn
 &+&\frac{2k-5}{12}\g^2\g^3\pa\g^1+\frac{k-2}{2}\g^{23}\pa\g^1
  -\frac{k}{3}\g^1\g^3\pa\g^2+\frac{2k+5}{12}\g^1\g^2\pa\g^3\nn
 &-&\frac{k+2}{2}\g^{12}\pa\g^3
  +\frac{k-1}{2}\g^3\pa\g^{12}-\frac{k+1}{2}\g^1\pa\g^{23}+k\pa\g^{123}
\label{FFRsl22}
\eea

\section{Generators of Asymmetric $N=4$ SCA}

For completeness and reference, below are listed explicit free field
realizations of the integrands ${\cal A}_n$ of the generators $A_n$ 
(\ref{agen}) of the asymmetric $N=4$ SCA
\ben
 A=\oint\dz{\cal A}_n(z)
\een
The realizations are obtained by inserting the results of Appendix B
in the general expressions for the generators provided in Section 3.
The realizations of $\phi_{-\al_2}$, $\phi_{-(\al_2+\al_3)}$ and $S$
may be obtained by working out explicitly the relevant (anti-)commutators
of the asymmetric $N=4$ SCA (\ref{N4a}).

The Virasoro generator is realized as
\bea
 {\cal L}_n&=&-(\g^1)^{n+1}\beta_1-\frac{n+1}{2}(\g^1)^{n-1}
  \left(\frac{n-2}{2}\g^1\g^2+n\g^{12}\right)\beta_2\nn
 &+&(\g^1)^n\left(\frac{n(n-3)}{8}\g^1\g^2+\frac{n^2-n-2}{4}\g^{12}\right)
  \beta_{12}\nn
 &+&\frac{n+1}{2}(\g^1)^{n-1}\left(\frac{n}{12}\g^1\g^2\g^3
  -\frac{n-2}{2}\g^1\g^{23}-n\g^{123}\right)\beta_{23}\nn
 &-&(\g^1)^n\left(\frac{n(n-1)}{24}\g^1\g^2\g^3-\frac{n(n-3)}{8}\g^1\g^{23}
  +\frac{n(n+1)}{24}\g^{12}\g^3\right.\nn
 &&\left.-\frac{(n+1)(n-2)}{4}\g^{123}\right)
  \beta_{123}+\hf(n+1)(\g^1)^n\sqrt{k}\pa\var_1
\eea
Here and throughout Appendix C we have taken advantage of the fact
that terms proportional to $n(\g^1)^{n-1}\pa\g^1$ vanish upon integration.
They are not included in the expressions for 
the integrands. The supercurrents are realized as 
\bea
 {\cal G}_{\al_2;n+1/2}&=&(\g^1)^n\left((n+1)\beta_2
  -\frac{n-1}{2}\g^1\beta_{12}-\frac{n+1}{2}\g^3\beta_{23}
  +\frac{2n-1}{6}\g^1\g^3\beta_{123}\right)\nn
 {\cal G}_{\al_2+\al_3;n+1/2}&=&(\g^1)^n\left((n+1)\beta_{23}
  -\frac{n-1}{2}\g^1\beta_{123}\right)
\eea
and
\bea
 \overline{{\cal G}}_{-\al_2;n-1/2}&=&-(\g^1)^n\left(\hf\g^1\g^2+\g^{12}
  \right)\beta_1-n(\g^1)^{n-1}\g^2\g^{12}\beta_2\nn
 &-&(\g^1)^{n-1}\left(\frac{2n-3}{6}\g^1\g^2\g^3
  -\frac{n-2}{2}\g^1\g^{23}+\hf n\g^{12}\g^3
  -n\g^{123}\right)\beta_3\nn
 &+&\frac{n-1}{2}(\g^1)^n\g^2\g^{12}\beta_{12}\nn
 &-&(\g^1)^{n-2}\left(\frac{n^2-2}{4}(\g^1)^2\g^2\g^{23}+
  \frac{n(n+2)}{12}\g^1\g^2\g^3\g^{12}\right.\nn
 &&\left.+\frac{n(n-1)}{2}\g^1\g^{12}\g^{23}
  +\frac{n^2}{2}\g^1\g^2\g^{123}+n(n-1)\g^{12}\g^{123}\right)\beta_{23}\nn
 &+&(\g^1)^{n-1}\left(\frac{n^2-4}{24}\g^1\g^2\g^{12}\g^3
  +\frac{3n^2-4n-4}{24}(\g^1)^2\g^2\g^{23}\right.\nn
 &&\left.+\frac{n(n-2)}{4}\g^1\g^{12}\g^{23}
  +\frac{(3n-4)n}{12}\g^1\g^2\g^{123}
  +\frac{n(n-2)}{2}\g^{12}\g^{123}\right)\beta_{123}\nn
 &+&n(\g^1)^{n-1}\left(\hf\g^1\g^2+\g^{12}\right)\sqrt{k}\pa\var_1\nn
 &+&(\g^1)^{n-1}\left(\frac{n-2}{2}\g^1\g^2+n\g^{12}\right)
  \sqrt{k}\pa\var_2\nn
 &+&\frac{(6kn-3n-8)n}{12}(\g^1)^{n-1}\g^2\pa\g^1
  +(k-1/2)(n^2-n)(\g^1)^{n-2}\g^{12}\pa\g^1\nn
 &+&\frac{(3n-6)k-n}{6}(\g^1)^n\pa\g^2
  +(k-1/2)n(\g^1)^{n-1}\pa\g^{12}\nn
 \overline{{\cal G}}_{-(\al_2+\al_3);n-1/2}&=&-(\g^1)^n\left(\frac{1}{6}
  \g^1\g^2\g^3+\hf\g^1\g^{23}+\hf\g^{12}\g^3+\g^{123}\right)\beta_1\nn
 &+&(\g^1)^{n-2}\left(\frac{n(n-7)}{12}\g^1\g^2\g^{12}\g^3
  +\frac{n^2+n-4}{4}(\g^1)^2\g^2\g^{23}\right.\nn
 &&\left.+\frac{n(n+1)}{2}\g^1\g^{12}\g^{23}+\frac{n(n-1)}{2}\g^1\g^2
  \g^{123}+n(n-1)\g^{12}\g^{123}\right)\beta_2\nn
 &-&(\g^1)^{n-1}\g^3\left(\frac{2n-3}{6}\g^1\g^2\g^3-\frac{n-2}{2}\g^1\g^{23}
  +\frac{n}{2}\g^{12}\g^3-n\g^{123}\right)\beta_3\nn
 &-&(\g^1)^{n-1}\left(\frac{n^2-9n+6}{24}\g^1\g^2\g^{12}\g^3
  +\frac{n^2-n-2}{8}(\g^1)^2\g^2\g^{23}\right.\nn
 &&\left.+\frac{n(n-1)}{4}\g^1\g^{12}\g^{23}
  +\frac{n^2-3n+2}{4}\g^1\g^2\g^{123}
  +\frac{n(n-3)}{2}\g^{12}\g^{123}\right)\beta_{12}\nn
 &-&(\g^1)^{n-2}\left(\frac{n(n+5)}{24}\g^1\g^2\g^{12}(\g^3)^2
  +\frac{n(n-1)}{4}\g^1\g^{12}\g^{23}\g^3\right.\nn
 &&+\frac{n(n-1)}{4}\g^1\g^2\g^3\g^{123}
  +\frac{(3n-1)n}{24}(\g^1)^2\g^2\g^{23}\g^3\nn
 &&\left.+\frac{n(n-1)}{2}
  \g^{12}\g^3\g^{123}+n\g^1\g^{23}\g^{123}\right)\beta_{23}\nn
 &+&(\g^1)^{n-1}\left(\frac{2n^2+7n-15}{72}\g^1\g^2\g^{12}(\g^3)^2
  +\frac{2n^2-n-3}{12}\g^1\g^{12}\g^{23}\g^3\right.\nn
 &&+\frac{2n^2-3n+1}{12}\g^1\g^2\g^3\g^{123}
  +\frac{2n^2-n-3}{24}(\g^1)^2\g^2\g^{23}\g^3\nn
 &&\left.+\frac{n(n-1)}{3}\g^{12}\g^3\g^{123}
  +\frac{n-1}{2}\g^1\g^{23}\g^{123}\right)\beta_{123}\nn
 &+&(\g^1)^{n-1}\left(\frac{n}{6}\g^1\g^2\g^3+\frac{n}{2}\g^1\g^{23}
  +\frac{n}{2}\g^{12}\g^3+n\g^{123}\right)\sqrt{k}\pa\var_1\nn
 &+&(\g^1)^{n-1}\left(\frac{n-3}{6}\g^1\g^2\g^3+\frac{n-2}{2}\g^1\g^{23}
  +\frac{n}{2}\g^{12}\g^3+n\g^{123}\right)\sqrt{k}\pa\var_2\nn
 &+&(\g^1)^{n-1}\left(\frac{2n-3}{6}\g^1\g^2\g^3-\frac{n-2}{2}\g^1\g^{23}
  +\frac{n}{2}\g^{12}\g^3-n\g^{123}\right)\sqrt{k}\pa\var_3\nn
 &+&(\g^1)^{n-2}\left(\frac{(4kn-3n-7)n}{24}\g^1\g^2\g^3
  +\frac{(3k-2)n(n-1)}{6}\g^{12}\g^3\right.\nn
 &&\left.+\frac{(2kn-n-3)n}{4}\g^1\g^{23}+
  (k-1/2)n(n-1)\g^{123}\right)\pa\g^1\nn
 &+&\frac{kn-n+1}{6}(\g^1)^n\g^3\pa\g^2
  +\frac{(k-1)n}{2}(\g^1)^{n-1}\g^3\pa\g^{12}\nn
 &-&(\g^1)^{n-1}\left(\frac{8kn-12k+n^2+11n-22}{24}\g^1\g^2
  +\frac{(6k+n+11)n}{12}\g^{12}\right)\pa\g^3\nn
 &+&\frac{kn-2k-1}{2}(\g^1)^n\pa\g^{23}+kn(\g^1)^{n-1}\pa\g^{123}
\eea
The affine $SL(2)$ current subalgebra is realized as
\bea
 \tilde{{\cal E}}_n&=&-(\g^1)^n\beta_3-(\g^1)^{n-1}\left(\frac{n+1}{2}\g^1\g^2
  +n\g^{12}\right)\beta_{23}\nn
 &+&(\g^1)^n\left(\frac{3n-1}{12}\g^1\g^2
  +\frac{n-1}{2}\g^{12}\right)\beta_{123}\nn
 \tilde{{\cal H}}_n&=&(\g^1)^{n-1}\left(\frac{n+2}{2}\g^1\g^2+n\g^{12}\right)
  \beta_2-2(\g^1)^n\g^3\beta_3\nn
 &-&(\g^1)^n\left(\frac{n}{4}\g^1\g^2+\frac{n-2}{2}\g^{12}\right)\beta_{12}\nn
 &-&(\g^1)^{n-1}\left(\frac{5n}{12}\g^1\g^2\g^3+\frac{n+2}{2}\g^1\g^{23}
  +n\g^{12}\g^3+n\g^{123}\right)\beta_{23}\nn
 &+&(\g^1)^n\left(\frac{n}{4}\g^1\g^2\g^3+\frac{n}{4}\g^1\g^{23}
  +\frac{7n}{12}\g^{12}\g^3+\frac{n-2}{2}\g^{123}\right)\beta_{123}\nn
 &+&(\g^1)^n\sqrt{k}\pa\var_3\nn
 \tilde{{\cal F}}_n&=&-(\g^1)^{n-1}\left(\frac{n+3}{6}\g^1\g^2\g^3
  +\frac{n+2}{2}\g^1\g^{23}
  +\frac{n}{2}\g^{12}\g^3+n\g^{123}\right)\beta_2\nn
 &+&(\g^1)^n(\g^3)^2\beta_3\nn
 &+&(\g^1)^n\left(\frac{n+1}{12}\g^1\g^2\g^3
  +\frac{n}{4}\g^1\g^{23}
  +\frac{n-2}{4}\g^{12}\g^3+\frac{n-2}{2}\g^{123}\right)\beta_{12}\nn
 &+&(\g^1)^{n-1}\g^3\left(\frac{n-3}{12}\g^1\g^2\g^3
  +\frac{n+2}{4}\g^1\g^{23}
  +\frac{n}{4}\g^{12}\g^3+\frac{n}{2}\g^{123}\right)\beta_{23}\nn
 &-&(\g^1)^{n}\g^3\left(\frac{n}{18}\g^1\g^2\g^3
  +\frac{2n+1}{12}\g^1\g^{23}
  +\frac{2n+3}{12}\g^{12}\g^3+\frac{2n-3}{6}\g^{123}\right)
  \beta_{123}\nn
 &-&(\g^1)^n\g^3\sqrt{k}\pa\var_3+\left(k+\frac{n}{12}+1\right)(\g^1)^n\pa
  \g^3
\eea
whereas the $U(1)$ generator is
\ben
 U_n=\oint\dz(\g^1)^n\sqrt{k}(-\pa\var_1-2\pa\var_2+\pa\var_3)
\label{Ureal}
\een
The remaining 2 fermionic spin 1/2 generators and the bosonic scalar are
\bea
 \phi_{-\al_2;n-1/2}&=&\oint\dz(\g^1)^{n-1}V_{-\al_2}^{\al_1}\sqrt{k}
  (-\pa\var_1-2\pa\var_2+\pa\var_3)\nn
 \phi_{-(\al_2+\al_3);n-1/2}&=&\oint\dz
  (\g^1)^{n-1}V_{-(\al_2+\al_3)}^{\al_1}\sqrt{k}
  (-\pa\var_1-2\pa\var_2+\pa\var_3)
\label{phireal}
\eea
and
\ben
 S_n=\oint\dz(\g^1)^{n-1}V_{-(\al_2+\al_3)}^{\al_1}
  V_{-\al_2}^{\al_1}\sqrt{k}(-\pa\var_1-2\pa\var_2+\pa\var_3)
\label{Sreal}
\een
According to (\ref{FFRsl22}), the polynomials $V_{-\al_2}^{\al_1}$ and 
$V_{-(\al_2+\al_3)}^{\al_1}$ are 
\bea
 V_{-\al_2}^{\al_1}&=&\hf\g^1\g^2+\g^{12}\nn
 V_{-(\al_2+\al_3)}^{\al_1}&=&\frac{1}{6}\g^1\g^2\g^3+\hf\g^1\g^{23}
  +\hf\g^{12}\g^3+\g^{123}
\label{V}
\eea

\end{document}